\documentclass[conference]{IEEEtran}
\usepackage[left=0.625in,right=0.625in,bottom=1.1in,top=0.75in]{geometry}
\makeatletter
\def\ps@headings{%
\def\@oddhead{\mbox{}\footnotesize\rightmark \hfil \thepage}%
\def\@evenhead{\footnotesize\thepage \hfil \leftmark\mbox{}}%
\def\@oddfoot{}%
\def\@evenfoot{}}
\makeatother
\ifCLASSINFOpdf
  \usepackage[pdftex]{graphicx}
  \DeclareGraphicsExtensions{.pdf,.jpeg,.png,.eps,.jpg}
\else
  \usepackage[dvips]{graphicx}
  \DeclareGraphicsExtensions{.eps}
\fi
\usepackage[cmex10]{amsmath}
\usepackage{amssymb,amsthm}
\usepackage{mathtools, cuted}
\usepackage{lipsum, color}

\usepackage[tight,footnotesize]{subfigure}
\newcommand{\argmin}{\operatornamewithlimits{arg\,min}}
\newcommand{\argmax}{\operatornamewithlimits{arg\,max}}
\newcommand{\abs}[1]{\left\lvert{#1}\right\rvert}
\newcommand{\norm}[1]{\left\lVert{#1}\right\rVert}

\begin{document}
%
\title{On the Security of Angle of Arrival Estimation}
%
%
%
%
\author{\IEEEauthorblockN{Amr Abdelaziz, C. Emre Koksal and Hesham El Gamal} \IEEEauthorblockA{Department of Electrical and Computer Engineering\\
The Ohio State University\\
Columbus, Ohio 43201\\
}}
\maketitle

\begin{abstract}
Angle of Arrival ({AoA}) estimation has found its way to a wide range of applications. Much attention have been paid to study different techniques for AoA estimation and its applications for jamming suppression, however, security vulnerability issues of AoA estimation itself under hostile activity have not been paid the same attention. In this paper, the problem of AoA estimation in Rician flat fading channel under jamming condition is investigated. We consider the scenario in which a receiver with multiple antenna is trying to estimate the {AoA} of the specular line of sight ({LOS}) component of signal received from a given single antenna transmitter using a predefined training sequence. A jammer equipped with multiple antennas is trying to interrupt the {AoA} estimation phase by sending an arbitrary signal. We derive the optimal jammer and receiver strategies in various scenarios based on the knowledge of the opponent strategies and the available information about the communication channel. In all scenarios, we derive the optimal jammer signal design as well as its optimal power allocation policy. The results show the optimality of the training based Maximum Likelihood ({ML}) {AoA} estimator in case of randomly generated jamming signal. We also show that, the optimal jammer strategy is to emit a signal identical to the predefined training sequence turning the estimation process into a highest power competition scenario in which the detected AoA is the one for the transmitting entity of higher power. The obtained results are supported by the provided computer simulation.

\end{abstract}


%
\IEEEpeerreviewmaketitle

\section{Introduction}
AoA estimation is one of the most important applications of array signal processing. It has found its way to a wide range of applications in military, navigation, radar, law enforcement and some commercial communication systems. Anti-jamming is one of the most interesting applications of AoA estimation. Recently in \cite{antijammin_aoa}, an anti-jamming mechanism for receivers operating in a cognitive radio network using AoA estimation combined with adaptive beamforming is presented. The use of AoA estimation in wireless physical layer authentication is another promising application. In \cite{securearray}, Xiong and Jamieson proposed a novel signal processing technique leverage multi-antenna receiver to profile the directions at which a certain transmission arrives, using this \textbf{AoA} information to construct highly sensitive signatures that with very high probability uniquely identify each client. Despite the fact that much attention paid to the enhancement of AoA estimation and the promising results reported about its various applications, performance measure on AoA estimation under jamming attacks did not receive the same attention. To that end, we develop a framework in which we evaluate the optimal attack strategies to degrade the AoA estimation performance. Subsequently, we find the optimal estimator under jamming attack and evaluate its performance. We clearly identify the conditions under which the attacker successfully manages to derail the estimator and the cases in which the attack can be overcome by the legitimate receiver.
\par
In the general problem of AoA estimation, various signal models have been investigated in literature based on the number of estimated signal sources, modeling of noise process, cooperation between emitting and receiving nodes and implications of multipath environment. In a non cooperative environment, where the emitted signal is unknown, the subspace techniques such as multiple signal classifier {MUSIC} \cite{MUSIC} and estimation of signal parameters via rotational invariant technique {ESPRIT} \cite{ESPIRIT} were developed with no knowledge required about the emitted signal. Performance analysis of conditional and unconditional {ML-AoA} estimator have been presented in \cite{COND_ML}. There, conditional {ML} estimator model the unknown emitted signal as a random process where it is modeled as a deterministic unknown parameter in the  unconditional {ML} estimator. A crucial requirement for all the aforementioned {AoA} estimation techniques in a non cooperative regime that the number of signals that can be estimated has to be less than the receiver array size. However, in a hostile jamming environment or in a varying interference conditions this condition will fail if the rank of the received signal plus jamming/interference exceeds the array size. Meanwhile, in a cooperative environment, where transmitters and receivers share a predefined signal waveform, it was shown in \cite{AOA_KNOWN} that exploitation of the known signal waveform enables a successfull estimation of number of sources greater than the array size using ML estimator. Despite of this considerable advantage, a huge computational complexity of the ML estimator is considered a major disadvantage as it requires a $K$-dimensional search algorithm for estimating AoAs of $K$ incident signals. In \cite{AOA_KNOWN} this problem have been resolved by the so called decoupled ML (DEML) estimator which turns the $K$-dimensional search problem into $K$ one-dimensional search problem. Another advantage for the work therein that the results were obtained for an arbitrary interference covariance matrix.
\par
In this work, we explore the effect of Rician fading environment on {AoA} estimation in the presence of a hostile jamming activity. In Rician fading environment, the received signal can be decomposed into two components; one is the specular component results from the {LOS} path and the other is the diffuse component due to multipath reflections, or generally the non-line of sight component ({NLOS}). The {LOS} component can be considered fixed while the {NLOS} component can be best described as a Rayleigh fading channel. For {AoA} estimation purposes, only the specular {LOS} component is considered as a signal of interest ({SOI}) while the diffuse {NLOS} component is considered as an undesired interference. This fact turns the {AoA} estimation process into a challenging task due to the inherit correlation between specular and diffuse component carrying the same signal waveform. 
\par
Despite the results for {AoA} estimation for an arbitrary interference covariance matrix, under jamming conditions these results doesn't hold directly. That is because, in contrast to arbitrary interference, the jamming process is an intended hostile activity that may have the ability to adopt, maneuver or change strategies to achieve the intended goal. Hence, based on the capabilities, availability of information about the target receiver design and {CSI}, we analyze both optimal jammer and, on the other side, receiver strategies in a game theoretic approach. 
 


\section{System Model and Problem Statement}
\label{sec:model}
In the rest of this paper we use boldface uppercase letters for random vectors/matrices, uppercase letters for their realizations, bold face lowercase letters for deterministic vectors and lowercase letters for its elements. While, $(.)^{*}$ denotes conjugate of complex number, $(.)^{\dagger}$ denotes conjugate transpose, $\mathbf{I}_N$ denotes identity matrix of size $N$, $\mathbf{tr}(.)$ denotes matrix trace operator, $\textbf{var}(.)$ denotes variance of random variable, $\det(.)$ denotes matrix determinant operator and $\mathbf{1}_{m \times n}$ denotes a $m \times n$ matrix of all 1's. 
\subsection{System model}
\label{subsec:sysmodel}
\begin{figure}[!tb]
\centering
\includegraphics[width=3.4 in, height = 2.4 in]{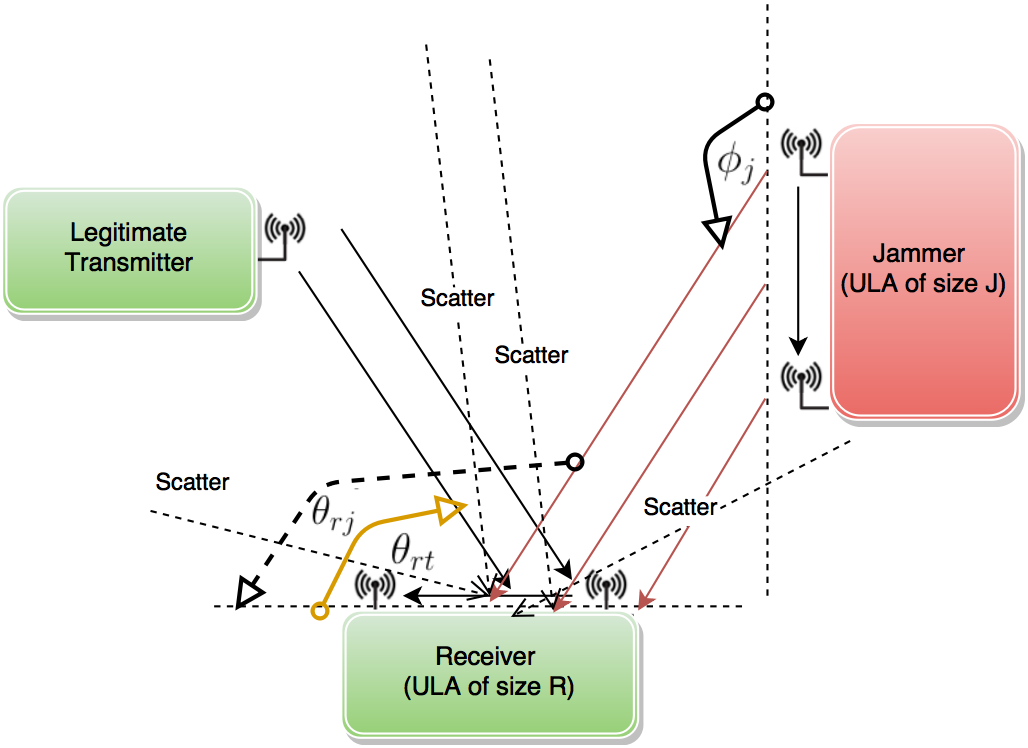}
\caption{System Model
\label{fig:sysmodel}}
\end{figure}

As depicted in Fig.\ref{fig:sysmodel}, we consider the scenario where a mobile receiver is trying to estimate the {AoA} of the the signal emitted from a certain fixed transmitter equipped with a single antenna in the presence of a multiple antenna jammer. The transmitter sends a predefined training sequence, $\mathbf{x}_{t} \in \mathbb{C}^L$, from which the {AoA} will be estimated.
The receiver is equipped with a uniform linear array (ULA) antenna that consists of $n_r$ elements placed along a linear array with neighboring antennas spaced at a distance $d_r$. On the other side, the jammer is equipped with a {ULA} antenna of size $n_j$. We assume a narrowband system under flat fading with a single significant channel tap. The discrete baseband equivalent channel for the signal received by the legitimate receiver at the $l^{th}$ time slot can be expressed as:
\begin{align}
\label{eq:inout} 
 \mathbf{Y}[l] &= {\mathbf{H}_t}[l] \mathbf{x}_{t}[l] + {\mathbf{H}_j}[l]{\mathbf{X}_j}[l] + \mathbf{N}[l],
\end{align}
where $\mathbf{x}_{t}[l] \in \mathbb{C}$ is the $l^{th}$ symbol of the predefined training sequence. The training sequence $\mathbf{x}_t =[x_{t1},..,x_{tL}]^T$ is constrained by both an instantaneous maximum power constraint $\abs{\mathbf{x}_{t}[l]}^2 \leq P_t^{\text{max}}$ and a sum power constraint $\mathbf{tr}\left(\mathbf{x}_t\mathbf{x}_t^{\dagger}\right) \leq P_t^{\text{tot}}$. The jammer signal ${\mathbf{X}_j}[l]$ is $n_j$ dimensional and it satisfies a total power constraint $\mathbf{tr}(\mathbf{Q}_j[l]) \leq \mathbf{P}_j$., where $\mathbf{Q}_j[l] = \mathbf{X}_j[l]\mathbf{X}_j^{\dagger}[l]$. Also, ${\mathbf{H}_t}[l] \in {\mathbb{C}^{n_r \times 1}}$ is the channel coefficients vector between transmitter and receiver, ${\mathbf{H}_j}[l] \in {\mathbb{C}^{n_r \times n_j}}$ is the channel coefficients matrix between jammer and receiver at the $l^{th}$ time index and $\mathbf{N}[l] \in \mathbb{C}^{n_r\times1}$ is an independent zero mean circular symmetric complex random vector, $\mathbf{N}\sim \mathcal{CN}(0,R_{n})$ where $R_{n} = \sigma_n^2 \mathbf{I}_{n_r}$. Thus, we can define the Signal to Jamming and Noise Ratio (SJNR) as follows:
\begin{align}
\text{SJNR} = \dfrac{\norm{{\mathbf{H}_t}[l] \mathbf{x}_{t}[l]}^2}{\norm{{\mathbf{H}_j}[l]{\mathbf{X}_j}[l]}^2+\sigma_n^2}. \nonumber
\end{align}


While the results we will drive are valid for all stationary and ergodic channel models, we will illustrate our examples on the Rician fading channel. Therefore, we next give some basics of the Rician fading channels. In Rician fading model, the received signal can be decomposed into two components; one is the specular component results from the {LOS} path and the other is the diffuse component due to multipath reflections. or generally the non-line of sight component ({NLOS}). The {LOS} component can be considered fixed while the {NLOS} component can be best described as a Rayleigh fading channel. Since both transmitter and jammer channels are Rician, we give the channel description for both transmitter and jammer channels with the subscripts $t,j$ are dropped. We consider a general transmitter and receiver with $N_t$ and $N_r$ array sizes.
\begin{align}
\label{eq:ric_decomp}
\mathbf{H} = \mathbf{H}^{\footnotesize{\text{LOS}}} + \mathbf{H}^{\footnotesize{\text{NLOS}}}, 
\end{align}
where $\mathbf{H}^{\footnotesize{\text{LOS}}}$ and $\mathbf{H}^{\footnotesize{\text{NLOS}}}$ represents the LOS and NLOS components respectively and
\begin{align}
\label{eq:ric_decomp_det}
\mathbf{H}^{\footnotesize{\text{LOS}}} &= \sqrt{\dfrac{k}{1+k}} \left(\dfrac{1}{\sqrt{2}}+\dfrac{j}{\sqrt{2}}\right) \hat{\mathbf{\Psi}} \nonumber \\
\hat{\mathbf{\Psi}} &= \mathbf{a}_r(\theta)\mathbf{a}_t^{\dagger}(\phi) \nonumber \\
\mathbf{H}^{\footnotesize{\text{NLOS}}}  &= \sqrt{\dfrac{1}{2(1+k)}}\hat{\mathbf{H}}, 
\end{align}
where $k$ is the Ricean factor, $\mathbf{a}_r(\theta)$ and $\mathbf{a}_t(\phi)$ are the antenna array steering vectors at receiver and transmitter respectively, $\theta$ and $\phi$ are the {AoA} and the angle of departure ({AoD}) of the transmitted signal respectively, as shown in Fig.\ref{fig:sysmodel}. In case of single antenna transmitter, we consider $\mathbf{a}_t(\phi)=1$. There are no prior distribution assumed for $\theta$ and $\phi$. Therefore, the associated estimation problem is non-Bayesian. For the {ULA} configuration, the entries of the steering vectors are given by 
\begin{align}
\label{eq:steering}
  &\mathbf{a}(\theta) =
 \begin{bmatrix}
     1 &  z & z^2 & \dots & z^{N-1}
\end{bmatrix}^T \nonumber \\
&z = e^{-j2\pi\dfrac{d\sin(\theta)}{\lambda}},
 \end{align}
where $\lambda$,$d$, and $N$ are the wavelength of the center frequency of the transmitted signal, array elements spacing and size respectively. While $\hat{\mathbf{H}} \sim \mathcal{CN}(\mathbf{0},\mathbf{I}_{N_r \times N_t})$ represents the channel coefficients matrix for the NLOS signal component. We parametrize the contribution of the NLOS and LOS components to the signal with $\sigma = \sqrt{{1}/{2(1+k)}}$, $\mu = \sqrt{{k}/{1+k}}$, respectively and choose $\mu^2 + 2\sigma^2 = 1$ for simplicity.
It worth mentioning that, AWGN and Rayleigh fading channels are in fact a limiting cases of the Rician fading channel.
\subsection{Attack Model}
\label{subsec:attack}
In this paper we consider two different jamming scenarios, \textbf{signal-unaware} and \textbf{signal-aware} jamming attacks. By signal-unaware jamming we refer to the scenario in which the jammer has no knowledge of the estimator, $\hat{\theta}(\cdot)$, the training sequence $\mathbf{x}_t$ and the realization of the channels, ${{H}_t}[l]$ and ${H}_j[l]$. Signal-unaware jammer only knows  the channels distributions. Meanwhile, the signal-aware jammer knows the estimator, the training sequence, and has perfect knowledge of the realization of both channels.
\subsection{Problem Statement}
\label{subsec:problem}

With the above setup we consider the problem in which the receiver with the knowledge of the predefined training pattern, $\mathbf{x}_t$, has the objective to choose an estimator $\hat{\theta}(\mathbf{Y})$ that has the minimum possible estimator variance under all possible jamming strategies for the choice of $\mathbf{X}_j[l],\ l\in \{1,\ldots , L\}$,  subject to its power constraint. On the other side, by an appropriate design of its transmitted signal, the jammer seeks to maximize the receiver estimator variance over all possible receiver strategies subject to its power constraint. Hence, we can formulate our problem as follows:  
  \begin{align}
  \label{eq:rx_obj}
  &\min_{\substack{\hat{\theta}(\cdot),\mathbf{x}_t \\ \abs{\mathbf{x}_{t}[l]}^2 \leq P_t^{\text{max}}\\ \mathbf{tr}({\mathbf{X}}_t{\mathbf{X}}_t^{\dagger}) \leq P_t^{\text{tot}}}} \;\; \max_{\substack{\mathbf{X}_j[l] \\\mathbf{tr}(\mathbf{Q}_j[l]) \leq \mathbf{P}_j\\ \forall l \in {1,..,L}}} \;\; \textbf{var}\left(\hat{\theta}\left(\mathbf{Y}\right)\right). 
 \end{align}  
\section{Basic Limits Under No Attack}
\label{sec:crb}
In this section, we evaluate the basic limits of AoA estimation performance in the absence of the attacker. In particular, we find a lower bound to the solution of the following problem:
\begin{align}
  \label{eq:rx_obj_nojam}
  &\min_{\substack{\hat{\theta}(\cdot),\mathbf{x}_t \\ \abs{\mathbf{x}_{t}[l]}^2 \leq P_t^{\text{max}}\\ \mathbf{tr}({\mathbf{X}}_t{\mathbf{X}}_t^{\dagger}) \leq P_t^{\text{tot}}}}  \;\; \textbf{var}\left(\hat{\theta}\left(\mathbf{Y}\right)\right). 
 \end{align} 
The solution of (\ref{eq:rx_obj_nojam}) is the CRB for AoA estimation by definition. To evaluate the CRB, we start by introducing
\begin{align}
\mathbf{Z}[l] = \mathbf{H}_t^{\footnotesize{\text{NLOS}}}[l]\mathbf{x}_{t}[l] + \mathbf{N}[l],
\end{align}
where $\mathbf{Z}$ incorporates all undesired interfering components of the received signal. Since the receiver objective is to estimate the AoA of the LOS component, the NLOS diffuse component is also considered as an undesired signal. Note that $\mathbf{Z}[l] \sim \mathcal{CN}\left(\mathbf{0}_{n_r \times 1},\mathbf{R}_z[l]\right)$. 
 Accordingly, the posterior distribution of the observation $\mathbf{Y}$ is given as follows:
\begin{flalign}
\label{eq:posterior_with_jam}
f_{\left(\mathbf{Y}/{H}_t^{\footnotesize{\text{LOS}}},\mathbf{x}_t,\mathbf{Z}\right)} (\mathbf{Y}) = \dfrac{1}{\prod_{i=1}^{L} \det(\pi\mathbf{R}_{z}[l])}  \nonumber 
\end{flalign}
\begin{align} 
 \times \exp \bigl\{-\dfrac{1}{L}\sum_{i=1}^{L}&(\mathbf{Y}[l]-{H}_t^{\footnotesize{\text{LOS}}}[l]{\mathbf{x}_t}[l])^{\dagger}\mathbf{R}_{z}[l]^{-1}  \nonumber \\
&(\mathbf{Y}[l]-{H}_t^{\footnotesize{\text{LOS}}}[l]{\mathbf{x}_t}[l])\bigr\}, 
\quad
\end{align}
which yields the following log-likelihood function
\begin{align}
\mathcal{L}(\mathbf{Y}) &= - \sum_{i=1}^{L} \ln \det(\pi\mathbf{R}_{z}[l]) \nonumber \\  
& - \mathbf{tr}\left(\mathbf{R}_{z}^{-1}(\mathbf{Y}-{H}_t^{\footnotesize{\text{LOS}}}{\mathbf{x}_t})(\mathbf{Y}-{H}_t^{\footnotesize{\text{LOS}}}{\mathbf{x}_t})^{\dagger}\right), 
\end{align}
where $\mathbf{R}_{z} = \dfrac{1}{L}\sum_{i=1}^{L}\mathbf{R}_{z}[l]$. Further, It can be shown that, the CRB of AoA estimation is given by
\begin{align}
 \label{eq:crb_stoica}
 \mathbf{CRB} &= \dfrac{1}{2}\left[\sum_{l=1}^{L}\mathbf{Re}\left(\mu_t\mathbf{x}_t^{\dagger}[l] \hat{\mathbf{D}}^{\dagger} \mathbf{G}(\theta)\hat{\mathbf{D}}\mathbf{x}_{t}[l]\mu_t\right)\right]^{-1}  \nonumber \\
 &= \dfrac{1+k_t}{2Lk_t P_t^{\text{max}} \hat{\mathbf{D}}^{\dagger} \mathbf{G}(\theta)\hat{\mathbf{D}}}, 
\end{align}
where
\begin{align}
&\hat{\mathbf{D}} = \mathbf{R}_{z}^{-1/2}\mathbf{D} \nonumber \\
&\mathbf{D}=\partial{\mathbf{a}}/\partial \theta \nonumber \\
&\mathbf{G}(\theta)  = [\mathbf{I} - {\mathbf{a}}({\mathbf{a}}^{\dagger}{\mathbf{a}})^{-1}{\mathbf{a}}^{\dagger}] \nonumber
\end{align}
where the dependence of $\mathbf{a}$ on $\theta$ was dropped for ease of notation. We note that, as $k_t \rightarrow \infty$, only LOS component is present and $\mathbf{R}_{z} \rightarrow \sigma_n^2\mathbf{I}_{n_r}$. Note that, this result is in agreement with that derived in \cite{AOA_KNOWN}. Also, one can show that (this was also discussed in \cite{AOA_KNOWN}) efficient estimator exists only asymptotically in the array size for any choice of $\mathbf{x}_t$ that satisfies the power constraint. Moreover, the {ML} estimator given by:
 \begin{align}
 \label{eq:ml_aoa}
 \hat{\theta}(\mathbf{Y}) &= \max_{\theta} \dfrac{\abs{\mathbf{a}^{\dagger}\mathbf{R}_{z}^{-1}\mathbf{B}}^2}{\mathbf{a}^{\dagger}\mathbf{R}_{z}^{-1}\mathbf{a}} \nonumber \\
 &= \min_{\theta} \mathbf{B}^{\dagger}\mathbf{R}_{z}^{-1/2} \mathbf{G}(\theta)  \mathbf{R}_{z}^{-1/2} \mathbf{B} 
\end{align} 
achieves the CRB with equality asymptotically in the large array size limit, where
\begin{align}
\label{eq:defs}
&\mathbf{B} = \mathbf{R}_{xy}^{\dagger} \mathbf{R}_{xx}^{-1} \;\;\; \in \mathbb{C}^{n_r \times 1} \\
&\mathbf{R}_{xy} = \dfrac{1}{L}\sum_{l=1}^{L}\mathbf{x}_{t}[l]\mathbf{Y}^{\dagger}[l] \;\;\;  \in \mathbb{C}^{1 \times n_r}  \nonumber \\
&\mathbf{R}_{xx} = \dfrac{1}{L}\sum_{l=1}^{L}\mathbf{x}_{t}[l]\mathbf{x}_{t}[l]^{*} \;\;\;  \in \mathbb{C}. \nonumber
\end{align} 
The results obtained in this section will be useful in the subsequent analysis in the rest of this paper.

\section{Optimal Jammer Strategies}
\label{sec:jam_objective} 
In this section we evaluate the optimal jammer strategies in the two different attack scenarios described in Section \ref{sec:model}. We start with the signal-unaware jamming scenario then we consider the worst case of jamming in which the jammer is aware with the receiver strategies. 

\subsection{Optimal Strategy for The Signal-Unaware Jammer}
\label{subsec:ign_jam}
In this subsection we consider the scenario where the only information available to the jammer is the statistical distribution of both channels. Here, the jammer has no knowledge on the estimator, $\hat{\theta}(\mathbf{Y})$, or the training sequence, $\mathbf{x}_t$. Hence, the attacker will target the optimal estimator performance, assuming that the receiver is using the (asymptotically) efficient estimator. Note that this attacker strategy also optimal for the attacker if the receiver is using the ML estimator. Thus, it will consider first the minimization problem in (\ref{eq:rx_obj_nojam}) whose solution is the {CRB} is given in Eq. (\ref{eq:crb_stoica}) for any arbitrary interference covariance matrix $\mathbf{R}_{z}$ which can be evaluated as
\begin{align}
\label{eq:eq_noise_jam}
\mathbf{R}_{z} &= \left(\dfrac{P_t^{\text{max}}}{1+k_t}+\sigma_n^2\right)\mathbf{I}_{n_r} + \mathbb{E}\bigl[{\mathbf{H}_j}{\mathbf{Q}_j}\mathbf{H}_j^{\dagger}\bigr]
\end{align}
(we derive this results in Appendix \ref{APP:APP_C}). Hence, the jammer objective is to find $\mathbf{X}_j$ and $\mathbf{Q}_j$ that both maximize the {CRB} expression and satisfy the power constraint. It can be formulated as follows:
\begin{align}
\label{eq:unaware_obj}
\mathbf{X}_j,\mathbf{Q}_j = \argmin_{\substack{\mathbf{X}_j,\mathbf{Q}_j \\ \mathbf{tr}(\mathbf{Q}_j)\leq \mathbf{P}_j}} {\mathbf{D}}^{\dagger} \mathbf{R}_{z}^{-1/2}\mathbf{G}(\theta)\mathbf{R}_{z}^{-1/2}{\mathbf{D}},  
\end{align} 
where the dependence on time slot index $l$ was dropped for ease of notation. We give the optimal jamming strategy as the solution of Eq. (\ref{eq:unaware_obj}) in the following theorem:
\par
\textbf{Theorem 1.} Given that statistical channel distribution is available at the jammer, the optimal jamming signal, $\mathbf{X}_j[l]$, that provides the solution of Eq. (\ref{eq:unaware_obj}) is a Gaussian vector with independent entries generated according to $\mathbf{X}_j[l] \sim \mathcal{CN}\left(0,\mathbf{Q}_j\right) \; \forall\; l \in \{1,2,..,L\}$, where $\mathbf{Q}_j$ is a diagonal matrix with
\begin{align}
 \footnotesize
\label{eq:diag_q_j}
{\mathbf{Q}}_j^{i,i} = \left\{
	\begin{array}{ll}
		\min\left\{\dfrac{\mathbf{P}_j}{n_j},\dfrac{k_j(1+k_j)}{n_r(1+n_jk_j)}\right\} n_j+ \left[\dfrac{\mathbf{P}_j}{n_j}-\dfrac{k_j(1+k_j)}{n_r(1+n_jk_j)}\right]^{+}  {i=1}  \\
		\left[\dfrac{\mathbf{P}_j}{n_j}-\dfrac{k_j(1+k_j)}{n_r(1+n_jk_j)}\right]^{+} \;\;\;\; {1 < i \leq n_j}
	\end{array}
 \right.,
\end{align}  
and $0 \leq k_j < \infty$ with $[x]^+ = \max\{0,x\}$.
\par 
\textbf{Proof:} The proof is given in Appendix \ref{APP:APP_A}.
\par
We note that, this result is in agreement with the optimal power allocation policy derived in \cite{rayl_capacity} for a transmitter aiming to maximize it's mutual information over a Rician MIMO channel. It also worth noting that, as the Rician factor, $k_j \rightarrow 0$, we notice that $\mathbf{Q}_j \rightarrow ({\mathbf{P}_j}/{n_j})\mathbf{I}_{n_j}$, which is the uniform power allocation policy. Thus, for the relatively small values of the Rician factor, $k_j$, the uniform power allocation policy is near optimal even if the channel distribution is unavailable at the jammer. The influence of using the uniform power allocation policy rather than the optimal one will be discussed in more details in Section \ref{sec:sim}.

\subsection{Optimal Strategy for The Signal-Aware Jammer}
\label{subsec:opt_jam_known}
In the previous section, we showed that with no knowledge about both receiver strategies and {CSI}, the optimal jammer strategy is to generate a Gaussian signal with power allocation policy as defined in (\ref{eq:diag_q_j}). In this section, we consider the case where both the training sequence, $\mathbf{x}_t$, and the estimator, $\hat{\theta}(\mathbf{Y})$, are known to the jammer. Also, perfect CSI about both channels in the form of channel realizations, $H_t[l]$ and $H_j[l] \;\;\forall l \in \{1,2,..,L\}$, are assumed to be known to a signal-aware jammer. In this case, the jammer objective is to find $\mathbf{X}_j$ and $\mathbf{Q}_j$ that both maximize the error of the estimator $\hat{\theta}(\mathbf{Y})$ and satisfy the power constraint. Since the ML estimator is an asymptotically efficient estimator, then, to maximize the ML estimator variance, a signal-aware jammer can consider maximizing the CRB based on its available information. Thus, the optimal jammer strategies are again the solution of Eq. (\ref{eq:unaware_obj}) taking into consideration the knowledge of the receiver strategies. We give the optimal jammer strategies in the following theorem:

\par
\textbf{Theorem 2.} Given that both the training sequence, $\mathbf{x}_t[l]$, and perfect CSI available at the jammer, the optimal jamming signal that provides the solution of Eq. (\ref{eq:unaware_obj}) can be found as:
  \begin{align}
 \footnotesize
\label{eq:diag_q_j_csi}
{\mathbf{Q}}_j^{i,i} =  \left\{
	\begin{array}{ll}
		\left(\mu-\lambda_{il}^{-1}\right)^{+}  & \mbox{if   } 1 < i \leq n \\
		0 \;\;\;\; &\mbox{if   } n < i \leq n_j 
	\end{array},
	\right.
\end{align} 
where $\lambda_{1l},\lambda_{2l},...,\lambda_{nl}$ are the eigenvalues of ${H}_j[l]{H}_j^{\dagger}[l]$ with $n=\min\{n_r,n_j\}$, $\mu$ is a constant chosen to satisfy the power constraint and
\begin{align}
\footnotesize
\label{eq:opt_X_j}
{\mathbf{X}}_j^{(i)}[l] = \left\{
	\begin{array}{ll}
		\sqrt{\left(\mu-\lambda_{il}^{-1}\right)^{+}}  \dfrac{\mathbf{x}_{t}[l]}{\abs{\mathbf{x}_{t}[l]}^2}   & \mbox{if  } 1 < i \leq n \\
		0 \;\;\;\; &\mbox{if  } n < i \leq n_j \
	\end{array}.
 \right.
\end{align}
\par
\textbf{Proof:} The proof is given in Appendix \ref{APP:APP_B}.
\par
Comparing the optimal jamming strategies in signal-unaware to that in the signal-aware jamming scenario we observe that, the knowledge of the training sequence $\mathbf{x}_t$ is a considerable advantage to the jammer. It grants the jammer the ability to design its signal aligned to the training sequence. Under this scenario for equal jammer and transmitter power, the ML spectrum output from the ML-AoA estimator shall be maximized in both the transmitter and jammer directions as can be seen in Fig. (\ref{fig:ML_EQ_POWER_NO_CSI}). Meanwhile, as the jammer power exceeds that of the transmitter, the ML spectrum is maximized in the jammer direction rather than the receiver one. Thus we conclude that, a signal aware jammer turns the AoA estimation process into a higher power competition in which the ML estimator outputs an estimate to the AoA of the transmitting entity of higher power. A more details about this scene are given in Section \ref{sec:sim}. We have studied the optimal jamming strategies in both attack scenarios, we  now turn our attention to the optimal receiver strategies in two scenarios, known and unknown jammer strategies.  

\section{Optimal AoA Estimator}
\label{sec:receiver_objective}
In this section, first evaluate the optimal estimator under the toy case in which the jammer strategies and CSI are known to the receiver. With the insights drawn, we solve the case in which the attacker strategy, as well as CSI is unknown to the receiver.
\subsection{Optimal AoA Estimator With Known Jammer Strategies}
Assuming the jammer strategies, $\mathbf{X}_j[l]\;\forall l$, and perfect CSI availability at the receiver, the noise, diffuse component and jamming covariance matrix is fully characterized at the receiver. That is because both channels realization and transmitted signals are known to the receiver. Thus, the receiver can substitute $\mathbf{R}_{z}$ directly into (\ref{eq:ml_aoa}) to form the ML estimator. Simulation results provided in Section \ref{sec:sim} show the superior performance of ML estimator with known jammer strategies. It shows that the knowledge of jammer strategies provides a considerable performance enhancement even for the case of signal-aware jammer. More discussion about these results will be provided in Section \ref{sec:sim}. The effect of unknown jamming strategies will be discussed in the next section. 

\subsection{Optimal AoA Estimator With Unknown Jammer Strategies}
In this section, the receiver is assumed to know only the statistical distribution of the channels. In Section \ref{subsec:opt_jam_known}, we showed that the worst case jamming strategy is to mate its transmitted signal to the target training sequence. In such scenario the interference covariance matrix can be evaluated as follows
\begin{align}
\label{eq:eq_noise_jam3}
\mathbf{R}_{z} &= \left(\dfrac{P_t^{\text{max}}}{1+k_t}+\sigma_n^2\right)\mathbf{I}_{n_r} +  {\mathbf{P}_{j}}\hat{\Upsilon}, 
\end{align}
where $\hat{\Upsilon} \in \mathbb{C}^{n_r \times n_r}$ is as defined in (\ref{eq:upsilon}) (the derivation of this result is given in Appendix \ref{APP:APP_C}). Direct substitution from (\ref{eq:eq_noise_jam3}) into (\ref{eq:crb_stoica}) and (\ref{eq:ml_aoa}) yields the CRB and the associated ML estimator solution for the optimization problem given in (\ref{eq:rx_obj}) for any choice of $\mathbf{x}_t$ that satisfies the power constraints.
\par
Simulation results provided shows that the ML estimator is inefficient against a signal-aware jammer despite the advantage of having the channel statistical distribution. As can be seen in Figures (\ref{fig:ML_EQ_POWER_STAT_CSI}), (\ref{fig:ML_DOUBLE_POWER_STAT_CSI}) and (\ref{fig:ML_4_POWER_STAT_CSI}), the normalized ML spectrum is maximized towards the direction of the transmitting entity of higher power direction even if statistical channel distribution is available at the receiver. 


\section{Simulation Results}
\label{sec:sim}
The simulation results provided in this section is based on the following simulation setup:
\begin{description}
\item[$\bullet$]\textbf{Array Size.} Both receiver and jammer are of array size $n_r=n_j=4$ while the transmitter has a single antenna.
  \item[$\bullet$]\textbf{Transmitter Signal.} The predefined training sequence $\mathbf{x}_t$, is generated from a zero mean, unit variance complex Gaussian random variable, fixed once chosen and shared to the receiver.
 \item[$\bullet$]\textbf{Jammer Signal.} In the Signal-Unaware scenario, the Jammer signal $\mathbf{X}_j[l]$, is generated from a zero mean, unit variance complex Gaussian random variable and scaled to satisfy the power constraint. Meanwhile, in the Signal-Aware scenario it is chosen identical to the training sequence.
  \item[$\bullet$]\textbf{Communication Channels.} The communication channels, $\mathbf{H}_t$ and $\mathbf{H}_j$, are generated according to Equations (\ref{eq:ric_decomp}) and (\ref{eq:ric_decomp_det}). The entries of the channel matrix of the Rayleigh part of the channel are generated from a zero mean, unit variance complex Gaussian random variable and then scaled each by the corresponding value of $\sigma$.
\end{description}
\subsection{Evaluation of The CRB}
\begin{figure}[!htbp]
\centering
\includegraphics[width=3.5 in, height = 2.4 in]{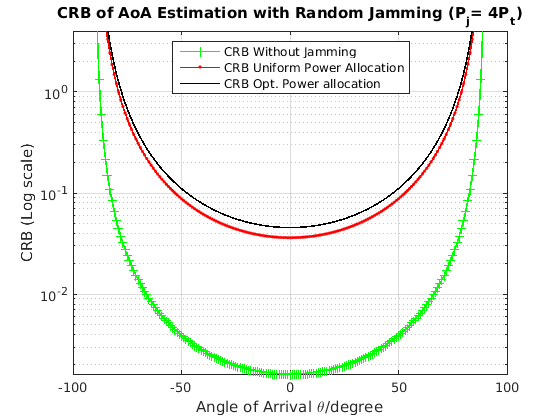}
\caption{CRB as a function of the {AoA} in log scale, $SNR = 15\text{dB}$, $P_j = P_t$, $k_j=10\text{dB}$}
\label{fig:CRB_Equal_Power_log}
\end{figure}

\begin{figure}[!htbp]
\centering
\includegraphics[width=3.5 in, height = 2.4 in]{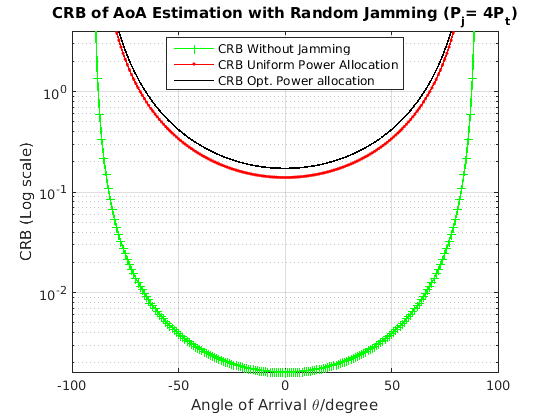}
\caption{CRB as a function of the {AoA} in log scale, $SNR = 15\text{dB}$, $P_j = 4P_t$, $k_j=10\text{dB}$}
\label{fig:CRB_4_Power_log}
\end{figure}

In Fig.(\ref{fig:CRB_Equal_Power_log}), the CRB is plotted as a function of the AoA, $\theta$, in logarithmic scale for a training sequence of length $64$ at a signal to noise ratio $(\text{SNR})=15\text{dB}$. While the jammer is emitting a random signal with power budget equal to that of the transmitter, which implies $(\text{SJNR}) < 0\text{dB}$. The figure provides comparison between the CRB for the jamming free environment to that for a jammer with uniform power allocation policy and also for a jammer uses the optimal power allocation policy derived in Eq. (\ref{eq:diag_q_j}). Note that, the uniform power allocation policy is optimal for $k_j=0$, hence, for relatively small $k_j$ values the uniform power allocation policy provides a perform close to optimal. To highlight the difference in their performance, we use $k_j=10\text{dB}$. As can be seen in Fig. (\ref{fig:CRB_Equal_Power_log}), the use of the optimal power allocation policy derived in Eq. (\ref{eq:diag_q_j}) has a superior performance compared to the uniform power allocation scheme in case of large values of the Rician factor $k_j$. Similar comparison is given in Fig. (\ref{fig:CRB_4_Power_log}) except that the jammer is emitting a signal identical to the training sequence with four times the power available at the transmitter. Note that, the use of the training sequence as a jamming signal did not affect the CRB. That is because the plotted CRB is for any unbiased estimator, however, using a jamming signal identical to the training sequence adds bias to the ML estimator. 
  
\par
  

\subsection{ML Estimator Performance under Jamming}
\begin{figure}[!htbp]
\centering
\includegraphics[width=3.5 in, height = 2.4 in]{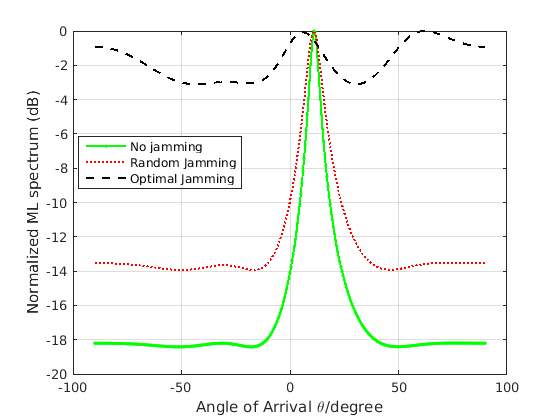}
\caption{ML Normalized Spectrum without CSI, $\theta_t = 12^{\circ}$, $\theta_j = 50^{\circ}$ and $P_j = P_t$ 
\label{fig:ML_EQ_POWER_NO_CSI}}
\end{figure}

\begin{figure}[!htbp]
\centering
\includegraphics[width=3.5 in, height = 2.4 in]{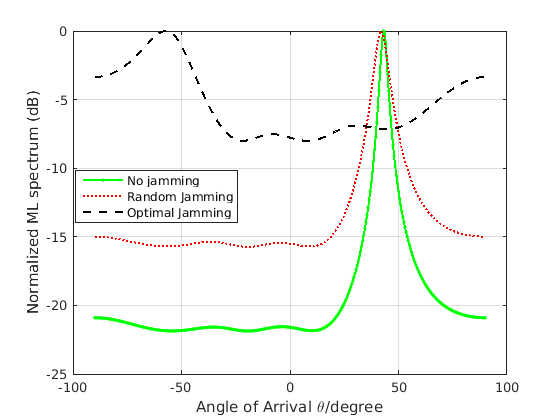}
\caption{ML Normalized Spectrum without CSI, $\theta_t = 43^{\circ}$, $\theta_j = -63^{\circ}$ and $P_j = 2 P_t$ 
\label{fig:ML_DOUBLE_POWER_NO_CSI}}
\end{figure}

\begin{figure}[!htbp]
\centering
\includegraphics[width=3.5 in, height = 2.4 in]{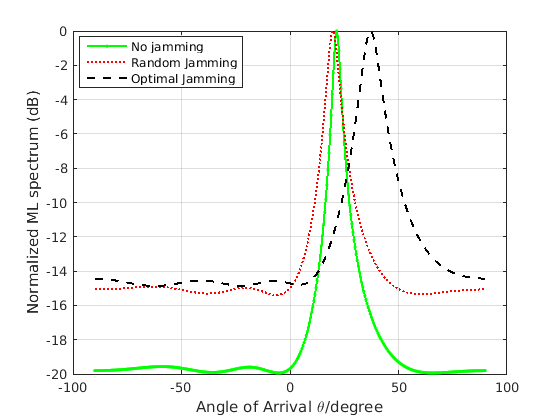}
\caption{ML Normalized Spectrum without CSI, $\theta_t = 23^{\circ}$, $\theta_j = 40^{\circ}$ and $P_j = 4 P_t$ 
\label{fig:ML_4_POWER_NO_CSI}}
\end{figure}

\begin{figure}[!htbp]
\centering
\includegraphics[width=3.5 in, height = 2.4 in]{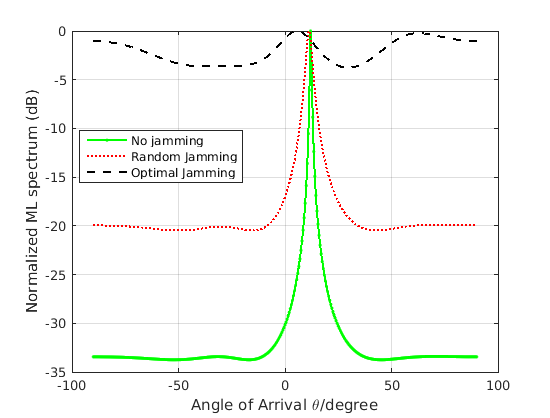}
\caption{ML Normalized Spectrum with Statistical Channel distribution, $\theta_t = 12^{\circ}$, $\theta_j = 50^{\circ}$ and $P_j = P_t$ 
\label{fig:ML_EQ_POWER_STAT_CSI}}
\end{figure}

\begin{figure}[!htbp]
\centering
\includegraphics[width=3.5 in, height = 2.4 in]{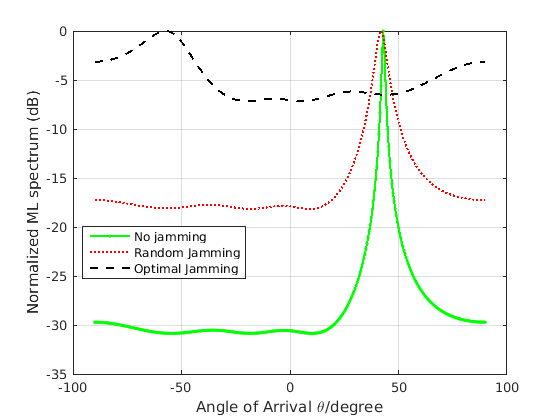}
\caption{ML Normalized Spectrum with Statistical Channel distribution, $\theta_t = 43^{\circ}$, $\theta_j = -63^{\circ}$ and $P_j = 2 P_t$
\label{fig:ML_DOUBLE_POWER_STAT_CSI}}
\end{figure}

\begin{figure}[!htbp]
\centering
\includegraphics[width=3.5 in, height = 2.4 in]{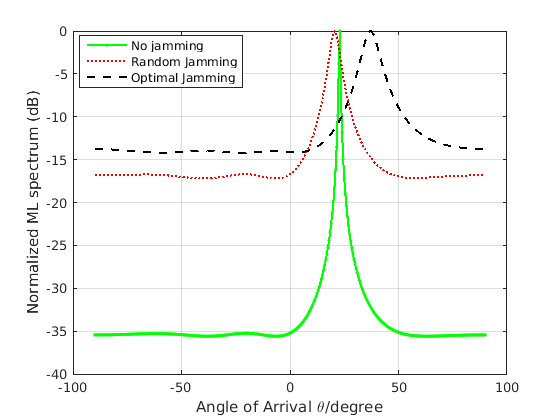}
\caption{ML Normalized Spectrum with Statistical Channel distribution, $\theta_t = 23^{\circ}$, $\theta_j = 40^{\circ}$ and $P_j = 4 P_t$
\label{fig:ML_4_POWER_STAT_CSI}}
\end{figure}

To evaluate the performance of the ML estimator under jamming conditions, we provide simulation results for different scenarios of jamming strategies as well as the amount of information available at the receiver. First we start by the case where no channel knowledge is available at the receiver. In Fig. (\ref{fig:ML_EQ_POWER_NO_CSI}, the transmitter is located  is at $12^{\circ}$, the jammer is located at $50^{\circ}$ and the jamming power equal to that of the transmitter. We notice that, For the signal-unaware jamming scenario where the optimal jamming signal is Gaussian generated as stated in Theorem 1, the {ML} output is maximized in the direction of the SOI as of the case of the jamming free model. As the jammer aligns its signal to the predefined training pattern, the {ML} estimator starts to bias towards the signal with higher power. This will be clear in Fig. (\ref{fig:ML_DOUBLE_POWER_NO_CSI}), where the transmitter is located  is at $43^{\circ}$, the jammer is located at $-63^{\circ}$ and the jamming power is twice that of the transmitter. While in Fig. (\ref{fig:ML_4_POWER_NO_CSI}) the transmitter is located  is at $23^{\circ}$ and the jammer is located at $40^{\circ}$ and the jamming power is four times the transmitter power. We see that in case of random jamming, the effect of jamming power appear as an increased estimator error. Meanwhile, as the jammer aligns its signal to the training sequence, the estimator is biased towards the jammer signal rather than transmitter one. The ML estimator exhibits the same behavior when only statistical channel distribution information is available at the receiver as seen in Figures (\ref{fig:ML_EQ_POWER_STAT_CSI}), (\ref{fig:ML_DOUBLE_POWER_STAT_CSI}) and (\ref{fig:ML_4_POWER_STAT_CSI}).
\par
\begin{figure}[!htbp]
\centering
\includegraphics[width=3.5 in, height = 2.4 in]{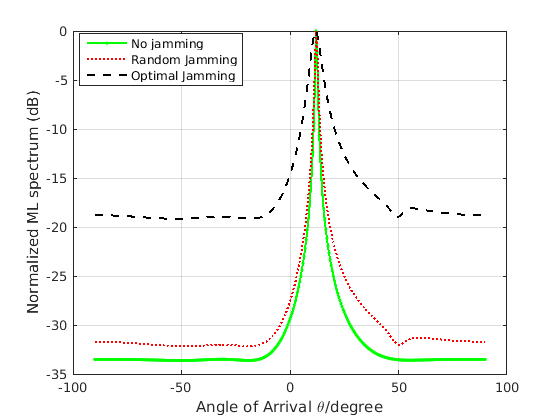}
\caption{ML Normalized Spectrum without CSI, $\theta_t = 12^{\circ}$, $\theta_j = 50^{\circ}$ and $P_j = P_t$ 
\label{fig:ML_EQ_POWER_CSI}}
\end{figure}
\begin{figure}[!htbp]
\centering
\includegraphics[width=3.5 in, height = 2.4 in]{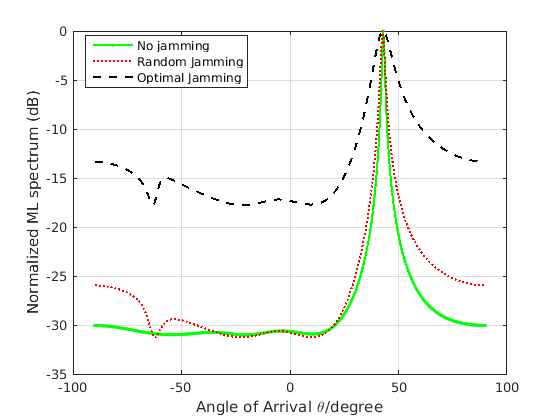}
\caption{ML Normalized Spectrum with perfect CSI, $\theta_t = 43^{\circ}$, $\theta_j = -63^{\circ}$ and $P_j = 2 P_t$
\label{fig:ML_DOUBLE_POWER_CSI}}
\end{figure}
\begin{figure}[!htbp]
\centering
\includegraphics[width=3.5 in, height = 2.4 in]{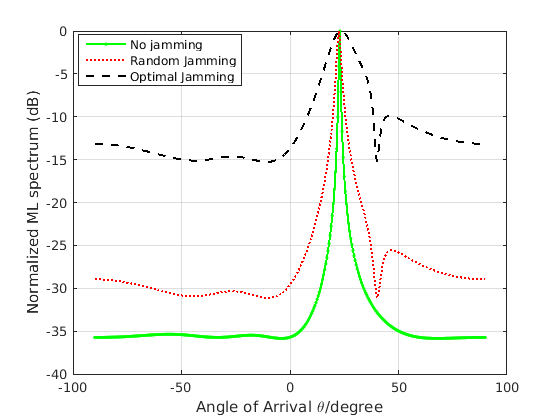}
\caption{ML Normalized Spectrum with perfect CSI, $\theta_t = 23^{\circ}$, $\theta_j = 40^{\circ}$ and $P_j = 4 P_t$
\label{fig:ML_4_POWER_CSI}}
\end{figure}

In case of perfect CSI availability about both channels at the receiver, we see in Figures (\ref{fig:ML_EQ_POWER_CSI}), (\ref{fig:ML_DOUBLE_POWER_CSI}) and (\ref{fig:ML_4_POWER_CSI}) that the ML estimator have a considerable performance even for an increased jamming power. 


\section{Conclusion}
\label{sec:conclusion}
The vulnerability of {AoA} estimation to hostile jamming activity is studied. we considered the problem of AoA estimation in Rician flat fading channel under jamming condition. We showed that in case of unknown receiver strategy, the optimal jamming signal is Gaussian. Moreover, if the jammer have the knowledge of the receiver strategy, its optimal signal design is to align its signal to the training pattern used for AoA estimation. Also, optimal power allocation policy based on the amount of information available at the jammer about the communication channel is provided. It is shown that, with {CSI} availability, the water filling power allocation policy is optimal while the uniform power allocation is optimal where no CSI available. From the receiver point of view, it was shown that the training based ML-AoA estimator has a superior performance in subject to the availability of perfect CSI. Computer simulation results are provided to support and demonstrate the obtained results. It showed the robustness of training based ML-AoA estimator under random jamming conditions. It also demonstrated the highest power competition in which the detected AoA is for the transmitting entity of higher power for a jammer with identical signal. In conclusion, the ML estimator works efficiently in all jamming scenarios in case of known jammer strategies. Meanwhile, in case of unknown jamming strategies, signal-aware jammer affects the ML performance to a large extent. 

\begin{appendices}
\section{Proof of Theorem 1}
\label{APP:APP_A}

We start from the  Signal-unaware jammer objective given in Eq. (\ref{eq:unaware_obj}) which is equivalent to
\begin{align}
\label{eq:obj_equiv}
\mathbf{X}_j,\mathbf{Q}_j &\stackrel{(a)}{=}  \argmin_{\substack{\mathbf{X}_j,\mathbf{Q}_j \\ \mathbf{tr}(\mathbf{Q}_j)\leq \mathbf{P}_j}} \mathbf{tr} \left({\mathbf{D}} \mathbf{R}_{z}^{-1/2}\mathbf{G}(\theta)\mathbf{R}_{z}^{-1/2}{\mathbf{D}}^{\dagger}\right) \nonumber \\
&\stackrel{(b)}{=} \argmin_{\substack{\mathbf{X}_j,\mathbf{Q}_j \\ \mathbf{tr}(\mathbf{Q}_j)\leq \mathbf{P}_j}} \mathbf{tr} \left(\mathbf{R}_{z}^{-1/2}{\mathbf{D}}^{\dagger}{\mathbf{D}} \mathbf{R}_{z}^{-1/2}\mathbf{G}(\theta)\right)  \nonumber \\ 
&\stackrel{(c)}{=} \argmin_{\substack{\mathbf{X}_j,\mathbf{Q}_j \\ \mathbf{tr}(\mathbf{Q}_j)\leq \mathbf{P}_j}} \mathbf{tr} \left(\Sigma\mathbf{R}_{z}^{-1}\mathbf{G}(\theta)\right),   
\end{align} 
where $(b)$ follows from the cyclic invariant property of the matrix trace operator and $(c)$ follows from 
\begin{align}
\Sigma = \mathbf{D}^{\dagger}\mathbf{D} = \left(\dfrac{2 \pi d_r\cos(\theta)}{\lambda}\right)^2 \sum_{i=0}^{n_r-1} i^2, 
\end{align}
which holds for the {ULA} configuration. Substituting (\ref{eq:eq_noise_jam}) and ignoring the constant terms as well as the terms that are independent to $\mathbf{X}_j$ and $\mathbf{Q}_j$ we get
\begin{align}
\label{eq:pow_aloc_no_csi}
\mathbf{X}_j,\mathbf{Q}_j &= \argmax_{\substack{\mathbf{X}_j,\mathbf{Q}_j \\ \mathbf{tr}(\mathbf{Q}_j)\leq \mathbf{P}_j}} \mathbf{tr} \left(\mathbb{E}\bigl[{\mathbf{H}_j}\mathbf{Q}_j\mathbf{H}_j^{\dagger}\bigr]\right) \nonumber \\
&\stackrel{(a)}{=} \argmax_{\substack{\mathbf{X}_j,\mathbf{Q}_j \\ \mathbf{tr}(\mathbf{Q}_j)\leq \mathbf{P}_j}} \mathbf{tr} \left(\mathbb{E}\bigl[\mathbf{H}_j^{\dagger}{\mathbf{H}_j}\mathbf{Q}_j\bigr]\right),
\end{align}
where $(a)$ follows from the cyclic invariant property of the matrix trace operator together with the fact that trace and expectation operators commute.
First, we need to evaluate the distribution of $\mathbf{X}_j$ that maximizes the above expression. We know that the entries of the channel matrix are Gaussian distributed as described in Section \ref{sec:model}. Hence, expanding the trace operator in the above expression and differentiating with respect to the distribution of $\mathbf{X}_j$ yields the entries of $\mathbf{X}_j$ should also be Gaussian distributed. Now, we need to show that the power allocation policy given in Theorem 1 is optimal.  


Based on the availability of channel statistical information, we evaluate 
\begin{align}
\label{eq:EHQH}
\mathbb{E}\left[\mathbf{H}_j^{\dagger}\mathbf{H}_j\right] = n_r\mathbf{\Upsilon},
\end{align}
where $\mathbf{\Upsilon}$ is $n_j\times n_j$ matrix and is given by:
\begin{align}
\label{eq:upsilon}
\mathbf{\Upsilon} = \dfrac{1}{1+k_j} \begin{bmatrix}
    1+k_j & k_j & \dots  & k_j \\
    k_j   & 1+k_j & \dots  & k_j \\
    \vdots & \vdots & \ddots & \vdots\\
    k_j & k_j & \dots  & 1+k_j
\end{bmatrix}.
\end{align}
For any value of $0 \leq k < \infty$, $\mathbf{\Upsilon}$ is non-singular, thus all it's eigenvalue are non-zero. The $n_j$ eigenvalues of $\mathbf{\Upsilon}$ are given by \cite{rician_mimo_capacity} :
\begin{align}
\label{eq:eigs}
\lambda_i = \left\{
	\begin{array}{ll}
		\dfrac{1+n_j k_j}{1+k_j}     & \mbox{if } i=1 \\
		\dfrac{1}{1+k_j}           & \mbox{if } 1 < i \leq n_j \
	\end{array}
 \right. \;\; 0 \leq k < \infty
\end{align}
We apply the eigenvalue decomposition to get $\mathbf{\Upsilon} = \mathbf{U}\mathbf{\Lambda}\mathbf{U}^{\dagger}$ where $\mathbf{\Lambda} \in \mathbb{C}^{n_j\times n_j}$ is a diagonal matrix having the eigenvalues in (\ref{eq:eigs}) as its diagonal entries, and $U \in \mathbb{C}^{n_j\times n_j}$ is a unitary matrix composed of the eigenvectors of $\mathbf{\Upsilon}$. Define $\tilde{\mathbf{Q}}_j = \mathbf{U}^{\dagger}\mathbf{Q}_j\mathbf{U}$, we get the following alternative expression for (\ref{eq:pow_aloc_no_csi})

\begin{align}
\label{eq:pow_aloc_csi}
\mathbf{X}_j,\mathbf{Q}_j &= \argmax_{\substack{\mathbf{X}_j, \mathbf{Q}_j \\ \mathbf{tr}(\mathbf{Q}_j)\leq \mathbf{P}_j}} \mathbf{tr} \left(\mathbb{E}\bigl[n_r\tilde{\mathbf{Q}}_j\mathbf{\Lambda}\bigr]\right).
\end{align}
The above expression is maximized for the choice of $\tilde{\mathbf{Q}}_j$ to be diagonal. The solution of the diagonal entries $\tilde{\mathbf{Q}}_j$ is found by the water filling algorithm as follows:
 \begin{align}
 \footnotesize
\label{eq:diag_q_j2}
\tilde{\mathbf{Q}}_j^{i,i} = \left\{
	\begin{array}{ll}
		\min\left\{\dfrac{\mathbf{P}_j}{n_j},\dfrac{k_j(1+k_j)}{n_r(1+n_jk_j)}\right\} n_j+ \left[\dfrac{\mathbf{P}_j}{n_j}-\dfrac{k_j(1+k_j)}{n_r(1+n_jk_j)}\right]^{+}     & \mbox{if } i=1 \\
		\left[\dfrac{\mathbf{P}_j}{n_j}-\dfrac{k_j(1+k_j)}{n_r(1+n_jk_j)}\right]^{+} \;\;\;\; \mbox{if } 1 < i \leq n_j \
	\end{array}
 \right.
\end{align} 
where $0 \leq k_j < \infty$ and $[x]^+ = \max\{0,x\}$. 
\par
We note that, the {CRB} evaluated for $\mathbf{Q}_j$ is the same as it evaluated for $\tilde{\mathbf{Q}}_j$. That is because $\mathbf{H}_j\mathbf{U}^{\dagger}$ has the same distribution as $\mathbf{H}_j$. Thus, the rsult of Theorem 1 follows immediately.
 
\section{Proof of Theorem 2}
\label{APP:APP_B}
\par
We take into consideration that, channel realizations of both transmitter and jammer channel are known to a signal-aware jammer, thus, we follow the proof of Theorem 1 except that we drop the expectation operator. We start by evaluating $\mathbf{R}_z$ in such case
\begin{align}
\label{eq:RZ_KNOWN}
\mathbf{R}_z &= \dfrac{1}{L} \sum_{l=1}^{L} \left({{H}_t^{{\text{NLOS}}}}[l]\mathbf{x}_{t}[l] + {H}_j[l] {X}_j[l] + \mathbf{N}[l]\right) \nonumber \\
&. \left({H}_t^{{\text{NLOS}}}[l] \mathbf{x}_{t}[l] +{H}_j[l]{{X}_j}[l]+ \mathbf{N}[l]\right)^{\dagger} \nonumber \\
&= \dfrac{1}{L} \sum_{l=1}^{L}{{H}_t^{{\text{NLOS}}}}[l]\mathbf{x}_{t}[l]\mathbf{x}_{t}^{\dagger}[l]({{H}_t^{{\text{NLOS}}}})^{\dagger}[l]\nonumber \\
&+{{H}_t^{{\text{NLOS}}}}[l]\mathbf{x}_{t}[l] {X}_j^{\dagger}[l]{H}_j^{\dagger}[l] \nonumber \\
&+{H}_j[l] {X}_j[l]\mathbf{x}_{t}^{\dagger}[l]({{H}_t^{{\text{NLOS}}}})^{\dagger}[l]\nonumber \\
&+{H}_j [l]{X}_j[l]{X}_j^{\dagger}[l]{H}_j^{\dagger}[l]+\sigma_n^2\mathbf{I}_{n_r} 
\end{align}

Starting from Eq. (\ref{eq:obj_equiv}) we substitute (\ref{eq:RZ_KNOWN}), ignore the constant terms as well as the terms that are independent to ${X}_j$ and ${Q}_j$ and drop the dependence on time slot index for ease of notation, we get
\begin{align}
\label{eq:pow_aloc_no_csi2}
{X}_j,{Q}_j &= \argmax_{\substack{{X}_j,{Q}_j \\ \mathbf{tr}({Q}_j)\leq \mathbf{P}_j}} \mathbf{tr} \left({{H}_t^{{\text{NLOS}}}}\mathbf{x}_{t} {X}_j^{\dagger}{H}_j^{\dagger}+{{H}_j}{Q}_j{H}_j^{\dagger}\right) \nonumber \\
&\stackrel{(a)}{=} \argmax_{\substack{{X}_j,{Q}_j \\ {tr}({Q}_j)\leq \mathbf{P}_j}} \mathbf{tr} \left({{H}_t^{{\text{NLOS}}}}\mathbf{x}_{t} {X}_j^{\dagger}{H}_j^{\dagger}\right)+\mathbf{tr} \left({H}_j^{\dagger}{{H}_j}{Q}_j\right),
\end{align}
where $(a)$ follows from the cyclic invariant property together with the linearity of the matrix trace operator. Without the power constraint, it is straightforward (Using Cauchy-Schwartz inequality) to see that the first trace expression in (\ref{eq:pow_aloc_no_csi2}) is maximized for $\mathbf{X}_j[l]$ whose entries are all equal to $\mathbf{x}_{t}[l]$, \textbf{i.e.}, the jammer aligns its signal to that of the transmitter signal. It remains to show that the power allocation policy stated in Theorem 2 is optimal. 
\par
Considering the maximization of the second trace expression, we apply the eigenvalue decomposition to get ${H}_j^{\dagger}{{H}_j} = {U}{\Lambda}{U}^{\dagger}$ where ${\Lambda} \in \mathbb{C}^{n_j\times n_j}$ is a diagonal matrix having the eigenvalues $\lambda_{1},\lambda_{2},...,\lambda_{n}$ as its first $n$ diagonal entries and the other $(n_j-n)$ entries are zeros where $n=\min\{n_r,n_j\}$, and $U \in \mathbb{C}^{n_j\times n_j}$ is a unitary matrix composed of the eigenvectors of ${H}_j^{\dagger}{{H}_j}$. Define $\tilde{{Q}}_j = {U}^{\dagger}{Q}{U}$, we get the following alternative expression for (\ref{eq:pow_aloc_no_csi2})
 \begin{align}
\label{eq:pow_aloc_csi2}
{X}_j,{Q}_j &= \argmax_{\substack{{X}_j, {Q}_j \\ \mathbf{tr}({Q}_j)\leq \mathbf{P}_j}} \mathbf{tr} \left(\tilde{{Q}}{\Lambda}\right).
\end{align}
The above expression is maximized for the choice of $\tilde{{Q}}_j$ to be diagonal. The solution of the diagonal entries $\tilde{{Q}}_j$ is found by the water filling algorithm and the theorem follows.
 
\section{Derivation of $\mathbf{R}_z$}
\label{APP:APP_C}
The interference, jamming and noise covariance matrix $\mathbf{R}_z$ can be evaluated as follows
\begin{align}
\mathbf{R}_{z} &\stackrel{(a)}{=}  \dfrac{1}{L} \sum_{l=1}^{L} \mathbb{E}\bigl[\left({\mathbf{H}_t^{\footnotesize{\text{NLOS}}}}[l]\mathbf{x}_{t}[l] + \mathbf{H}_j[l] \mathbf{X}_j[l] + \mathbf{N}[l]\right) \nonumber \\
&. \left(\mathbf{H}_t^{\footnotesize{\text{NLOS}}}[l] \mathbf{x}_{t}[l] +\mathbf{H}_j[l]{\mathbf{X}_j}[l]+ \mathbf{N}[l]\right)^{\dagger}\bigr] \nonumber \\
&\stackrel{(b)}{=} \dfrac{1}{L} \sum_{l=1}^{L} \mathbb{E}\bigl[{\mathbf{H}_t^{\footnotesize{\text{NLOS}}}}[l]\mathbf{x}_{t}[l] \mathbf{x}_{t}^{*}[l]({\mathbf{H}_t^{\footnotesize{\text{NLOS}}}})^{\dagger}[l]\bigr] \nonumber \\
&+ \mathbb{E}\bigl[{\mathbf{H}_j}[l] {\mathbf{X}_j}[l] \mathbf{X}_j^{\dagger}[l]\mathbf{H}_j^{\dagger}[l]\bigr] + \mathbb{E}\bigl[\mathbf{N}[l]\mathbf{N}^{\dagger}[l]\bigr] \nonumber \\
&\stackrel{(c)}{=} \left(\dfrac{P_t^{\text{max}}}{1+k_t}\right)\mathbf{I}_{n_r} + \dfrac{1}{L} \sum_{l=1}^{L} \mathbb{E}\bigl[{\mathbf{H}_j}[l]{\mathbf{Q}_j}[l]\mathbf{H}_j^{\dagger}[l]\bigr] + \sigma_n^2 \mathbf{I}_{n_r}, \nonumber
\end{align}
where $(a)$ is a direct substitution form the definition of the covariance matrix, in $(b)$ we used the fact that signals, channels and noise of both transmitter and jammer are independent. Also, we assumed the channels distribution to be time invariant. While in $(c)$, we used the distribution of the NLOS component of the received signal provided in Section \ref{sec:model}. Now using Eq. (\ref{eq:EHQH}),  the result given in Eq. (\ref{eq:eq_noise_jam3}) is immediate.

\end{appendices}
\normalsize



\bibliographystyle{IEEEtran}
\bibliography{IEEEabrv,References}
%
%
%

\end{document}